\documentclass[usenatbib]{mn2e}
\usepackage{natbibmnfix}
\usepackage{graphicx}
\usepackage{times}

\pdfoutput=1
\usepackage{epstopdf}
\usepackage{subfig}
\usepackage{float}

\newcommand{\rev}{}

\newcommand{\sunrise}{{\footnotesize SUNRISE}}
\newcommand{\gadthree}{{\footnotesize GADGET-3}}

\newcommand{\msun}{M$_{\odot}$}
\newcommand{\msunyr}{M$_{\odot}$ yr$^{-1}$}
\newcommand{\kms}{km s$^{-1}$}
\newcommand{\inv}{$^{-1}$}
\newcommand{\vk}{$v_{\rm k}$}
\newcommand{\vesc}{$v_{\rm esc}$}
\newcommand{\tmrg}{$t_{\rm mrg}$}

\newcommand{\fgas}{$f_{\rm gas}$}
\newcommand{\tagn}{$t_{\rm AGN}$}
\newcommand{\tobs}{$t_{\rm obs}$}
\newcommand{\hbeta}{H{$\beta$}}
\newcommand{\oiii}{[O{\rm\,III}]}

\newcommand{\lsun}{L$_{\odot}$}

\title[Constraints on the Nature of CID-42]{Constraints on the Nature of CID-42:\\ Recoil Kick or Supermassive Black Hole Pair?}

\author[Blecha et al.]{Laura Blecha$^1$\thanks{Email: lblecha@cfa.harvard.edu}, Francesca Civano$^1$, Martin Elvis$^1$, Abraham Loeb$^1$\\ $^1$ Harvard-Smithsonian Center for Astrophysics, 60 Garden St., Cambridge, MA 02138, USA}

\begin{document}
\maketitle

\begin{abstract}
The galaxy CXOC J100043.1+020637, also known as CID-42, is a highly unusual object. An apparent galaxy merger remnant, it displays signatures of both an inspiraling, kiloparsec-scale active galactic nucleus (AGN) pair and of a recoiling AGN with a kick velocity $\ga 1300$ \kms. Among recoiling AGN candidates, CID-42 alone has both spatial offsets (in optical and X-ray bands) {\em and} spectroscopic offsets. In order to constrain the relative likelihood of both scenarios, we develop models using hydrodynamic galaxy merger simulations coupled with radiative transfer calculations. Our gas-rich, major merger models are generally well matched to the galactic morphology and to the inferred stellar mass and star formation rate. We show that a recoiling supermassive black hole (SMBH) in CID-42 should be observable as an AGN at the time of observation. However, in order for the recoiling AGN to produce narrow-line emission, it must be observed shortly after the kick while it still inhabits a dense gaseous region, implying a large total kick velocity (\vk\ $\ga 2000$ \kms). For the dual AGN scenario, an unusually large broad-line offset is required, and the best match to the observed morphology requires a galaxy that is less luminous than CID-42. Further, the lack of X-ray emission from one of the two optical nuclei is not easily attributed to an intrinsically quiescent SMBH or to a Compton-thick galactic environment. While the current data do not allow either the recoiling or the dual AGN scenario for CID-42 to be excluded, our models highlight the most relevant parameters for distinguishing these possibilities with future observations. In particular, high-quality, spatially-resolved spectra that can pinpoint the origin of the broad and narrow line features will be critical for determining the nature of this unique source.
\end{abstract}


\section{Introduction}
\label{sec:intro}

Recent breakthroughs in numerical relativity have allowed for fully general-relativistic simulations of black hole (BH) mergers, including calculations of gravitational waveforms and kick velocities from gravitational-wave (GW) recoil \citep[e.g.,][]{pretor05,campan07a,baker08}. Surprisingly, these simulations have found that GW recoil kicks may be quite large, up to $\sim$ 5000 \kms\ \citep{campan07b,lousto11}. Thus, the merged supermassive black hole (SMBH) may be ejected from its host galaxy in some cases, and kick velocities of even a few percent of this maximum could be of astrophysical relevance. This discovery has prompted many studies of the possible electromagnetic (EM) signatures from recoiling SMBHs as well as observational searches for such objects. 

While prompt EM signatures of SMBH mergers are of great interest for identifying counterparts to eventual GW detections, long-lived signatures such as offset AGN provide the best chance of finding recoiling SMBHs using purely EM observations. An AGN ejected from the center of a galaxy can carry along its accretion disk as well as the broad-line region, which could produce an AGN spatially offset from its host galaxy or an AGN with offset broad emission lines \citep{madqua04, loeb07, bleloe08, kommer08b}. In either case the offset AGN lifetimes may be up to tens of Myr for a fairly wide range in kick speeds \citep{blecha11}. To date, several candidate recoiling SMBHs have been found with either spatial offsets \citep{batche10,jonker10} or kinematic offsets \citep{komoss08,shield09b,robins10}, though none have been confirmed \citep[for a review, see]{komoss12}. 

Recent years have also seen renewed energy in the search for inspiraling pairs of SMBHs. Systematic searches revealed that about 1\% of AGN have double-peaked narrow [OIII] lines, which could be a signature of SMBH orbital motion on kiloparsec scales \citep{comerf09a,liu10a,smith10}. Follow-up observations using higher-resolution imaging and slit or integral field unit (IFU) spectra have identified many strong dual AGN candidates \citep{liu10b,shen11,fu11a,fu12,mcgurk11,rosari11,comerf12}, and in two cases, evidence for dual compact X-ray \citep{comerf11a} and radio \citep{fu11b} sources, respectively, has been found. \citet{shen11} estimate that at least 10\% of double-peaked NL AGN may actually contain dual SMBHs.

\begin{figure*}
\centering
\subfloat[]{\includegraphics[width=0.32\textwidth]{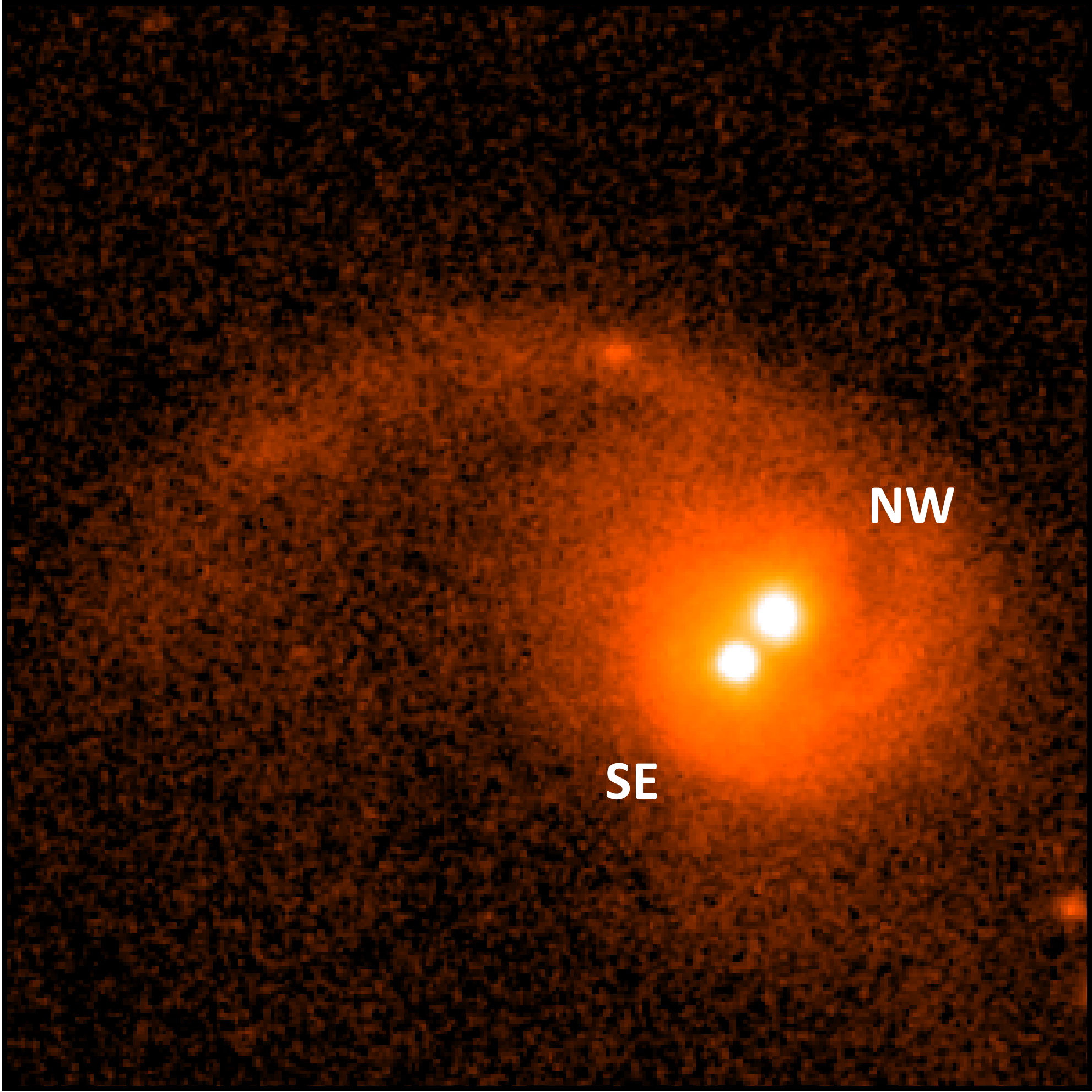}}
\hspace{5.5pt}
\subfloat[]{\includegraphics[width=0.32\textwidth]{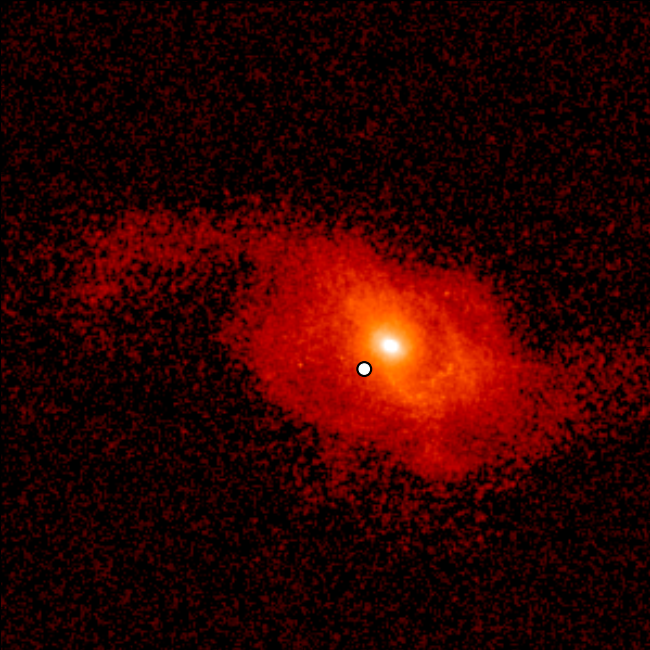}}
\hspace{6pt}
\subfloat[]{\includegraphics[width=0.32\textwidth]{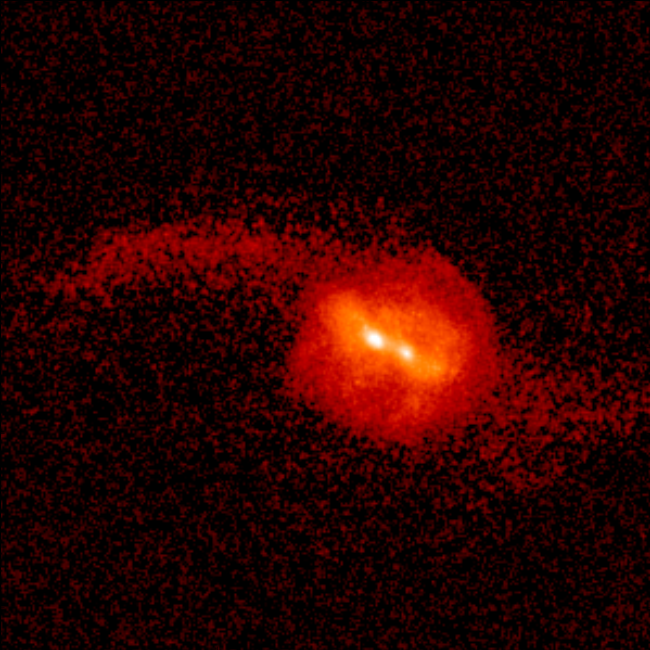}}
\caption{Comparison of 9" $\times$ 9" {\em HST}/ACS image of CID-42 ({\bf a}, Figure 3 from FC10) with simulated broadband (F814W) images of the recoiling ({\bf b}) and dual ({\bf c}) SMBH models. The orientation of the NW and SE optical brightness peaks is indicated in {\bf(a)}; north is up, east is to the left. The simulated images were generated with the radiative-transfer code SUNRISE. They have the same spatial and pixel scale (0.03"/pixel) as the {\em HST} image. They are shown at the time of observation (\tobs), i.e. when the spatial and velocity offsets of the SMBH match the observed values. The GW recoil model shown has \vk\ $= 1428$ \kms\ and \tobs\ $= t_{\rm mrg} + 5.9$ Myr, though the galactic morphology is essentially identical in the model with \vk\ $= 2470$ \kms\ and \tobs\ $= t_{\rm mrg} + 1.3$ Myr. The SMBH is offset from the stellar density peak by 2.5 kpc. In the dual SMBH model, \tobs $= t_{\rm mrg} - 50$ Myr, and the two BHs are separated by 2.5 kpc. 
\label{fig:stars} }
\end{figure*}

Among these new data and interpretations of SMBHs in merging galaxies, the galaxy CXOC J100043.1+020637 (also known as CID-42; $z = 0.359$) is a fascinating case. Discovered serendipitously by \citet[][hereafter JC09]{comerf09b} based on the striking morphology evident in its COSMOS {\em HST}/ACS image, it is an apparent face-on spiral with a prominent tidal tail and two well-resolved brightness peaks in its center, separated by 2.46 kpc (Figure \ref{fig:stars}a). The disturbed morphology strongly indicates a recent galaxy merger. DEIMOS slit spectroscopy revealed dual emission components in three different narrow lines (NLs) with a velocity separation of $\sim 150$ \kms\ and a spatial offset roughly consistent with the separation of the two brightness peaks in the image, albeit with low signal-to-noise (S/N, JC09). No evidence for broad lines (BLs) is apparent in this spectrum. Thus, JC09 conclude that CID-42 is a good candidate for a kpc-scale, Type 2 AGN pair.

Around the same time, CID-42 was also identified as an unusual object during visual inspection of the optical counterparts to {\em Chandra}-COSMOS sources \citep{elvis09,civano12a}. Additionally, a broad (\hbeta) line was detected in the SDSS spectrum of CID-42 \citep{shen11b} that was not seen in the DEIMOS spectrum. \citet[][hereafter FC10]{civano10} and \citet[][hereafter FC12]{civano12b} also found a prominent broad \hbeta\ line in IMACS/Magellan \citep{trump07,trump09} and VIMOS/VLT \citep[$z$COSMOS;][]{lilly07} spectra. More interestingly, the broad and narrow \hbeta\ components are {\em offset} by $\sim 1360$ \kms. The double-peaked narrow lines seen in the DEIMOS spectrum would be unresolved in these spectra, which have higher S/N but lower spectral resolution (R=700 on IMACS and VIMOS versus R=3000 on DEIMOS). There is also no measurable spatial offset between the broad and narrow emission components, though this may be attributable to seeing conditions (FC10). 

A BL/NL offset of 1360 \kms\ is much too large to be explained by orbital motion of dual AGN on kiloparsec scales. However, this is a signature expected from a {\em recoiling} AGN, which should retain only the most tightly bound gas and stars (including the BL region) and leave behind gas and stars at larger radii (including the NL region) \citep{merrit06}. Accordingly, FC10 proposed that CID-42 may contain a single, rapidly-recoiling SMBH rather than two inspiraling SMBHs. In this scenario, one of the optical brightness peaks (the NW source in the {\em HST} image, Figure \ref{fig:stars}a) corresponds to the central stellar cusp left behind by the recoiling SMBH, and the other (SE source) corresponds to the accretion luminosity of the recoiling AGN. A surface brightness decomposition of this image performed by FC10 is consistent with this picture, in that the SE source (the purported recoiling AGN) is best fit as a point source, while the NW source (the purported empty stellar cusp) is best fit with an extended Sersic component. 

Even more intriguing evidence as to the nature of CID-42 comes from High Resolution Chandra observations (FC12), which reveal that the X-ray emission detected in CID-42 arises from only the SE source, with a 3$\sigma$ upper limit of 4\% of the observed X-ray flux from the NW source. This finding supports the recoil scenario for CID-42, in which only one of the optical brightness peaks contains an AGN, but it does not rule out the possibility of a quiescent or Compton-thick AGN in the NW source.

Another possibility is a gravitational {\em slingshot} recoil resulting from a triple-SMBH interaction. This may occur if the binary SMBH inspiral time is long enough that the host galaxy undergoes a subsequent merger in the meantime, bringing a third SMBH into the system. We focus instead on GW recoil in our modeling, on the grounds that a triple-SMBH interaction is most likely in a gas-poor merger remnant, unlike CID-42. Nonetheless, a slingshot recoil cannot be ruled out by the observations, and this possibility is discussed further in \S~\ref{ssec:recoil}.

Because CID-42 displays signatures of both dual and recoiling AGN, and because it is the only recoil candidate with evidence for both spatial and kinematic offsets, this object is an ideal case study for theoretical models. \citet{blecha11} have conducted a large parameter study of recoiling SMBHs in hydrodynamic simulations of galaxy mergers, which allowed them to estimate observable lifetimes of spatially- and kinematically-offset recoiling AGN for a wide range of kick velocities and merging galaxy parameters. Here, we use hydrodynamic simulations to constrain the possible parameters of a recoiling or dual AGN in CID-42 and to gain insight about the relative plausibility of these scenarios. We provide models for both a GW-recoiling AGN and for an AGN pair, based on the methods of \citet{blecha11} and \citet{blecha12}. 

The details of our simulations are outlined in \S~\ref{ssec:sims}. We describe the properties of our best-fit recoiling SMBH simulations in \S~\ref{ssec:recoil_model}, and those of our best-fit dual SMBH simulation in \S~\ref{ssec:dual_model}. In \S~\ref{ssec:recoil}-\ref{ssec:dual}, we discuss the plausibility of the recoiling and dual AGN scenarios, and we briefly summarize in \S~\ref{ssec:summary}.

\section{Models}
\label{sec:model}

\subsection{Initial Conditions and Merger Simulations}
\label{ssec:sims}

To model the possible origins of CID-42, we begin by simulating a major galaxy merger using the smoothed-particle-hydrodynamics (SPH) code \gadthree\ \citep{spring05a}. Each progenitor galaxy contains dark matter, star, and gas particles, as well as an accreting, central SMBH. \gadthree\ also contains sub-resolution models for star formation and for supernova and AGN feedback.

We have designed simulations for both the GW recoil and dual SMBH scenarios that match as closely as possible the observed properties of CID-42. Specifically, the progenitor galaxies have a mass ratio of $q = 0.5$, and the primary galaxy has a baryonic (total) mass of $1.9 \times 10^{10}$ \msun\ ($4.5 \times 10^{11}$ \msun). Because star formation occurs throughout the simulation, the initial baryonic mass is chosen such that by the time the system would be observed, near the time of SMBH merger, the stellar mass matches that inferred for CID-42 ($\sim 2.5 \times 10^{10}$ \msun, FC10). The galaxies are disk-dominated with 10\% of the initial stellar mass in a bulge component. For the recoiling (dual) SMBH model, the initial gas fraction in the disk (\fgas) is 0.5 (0.4). The slightly lower gas content in the latter results in a more relaxed galaxy morphology prior to the SMBH merger, consistent with observations.

Each SMBH particle has an initial dynamical mass of $10^{-4}$ the total mass. The initial ``seed" mass (onto which accretion is calculated smoothly via the Bondi-Hoyle-Lyttleton formula) is $7 \times 10^5$ \msun\ for the recoil model and $10^6$ \msun\ for the dual SMBH model. The BH and star formation prescriptions are described in more detail in \citet{spring05b}. 

In the event of a high-velocity GW recoil kick, the Bondi accretion rate ($\propto \rho_{\rm gas} / v_{\rm BH}^3$) drops precipitously. Accretion onto the recoiling SMBH instead becomes dominated by the gas that remains bound to the SMBH. We adopt the ``ejected-disk" model described in \citet{blecha11}, in which the gas ejected along with the SMBH is modeled as an isolated accretion disk with a monotonically decreasing accretion rate normalized to the Bondi rate at the time of the kick. 

Assuming a minimum Eddington ratio $f_{\rm AGN} = \dot M_{\rm AGN}/\dot M_{\rm Edd}$ for the recoiling SMBH to be observable as an AGN, then the observable lifetime of the SMBH is given by
\begin{equation}
t_{\rm AGN} = {3 \over 16} \:\: {M_{\rm BH} \over \dot M_{\rm Edd} } \:\: {f_{\rm M_{\rm BH},ej} \over f_{\rm Edd,mrg} } \left [ \left ( { f_{\rm AGN} \over f_{\rm Edd,mrg} } \right )^{-16/19} - \, 1 \right ],
\label{eqn:tagn}
\end{equation}
where $f_{\rm M_{\rm BH},ej}$ is the initial mass of the ejected disk and $f_{\rm Edd,mrg}$ is the Eddington ratio at the time of SMBH merger. ($M_{\rm BH}$ and $\dot M_{\rm Edd}$ are time-dependent but change negligibly over the timescales with which we are concerned.) 

\begin{figure}
\centering
\includegraphics[width=0.49\textwidth]{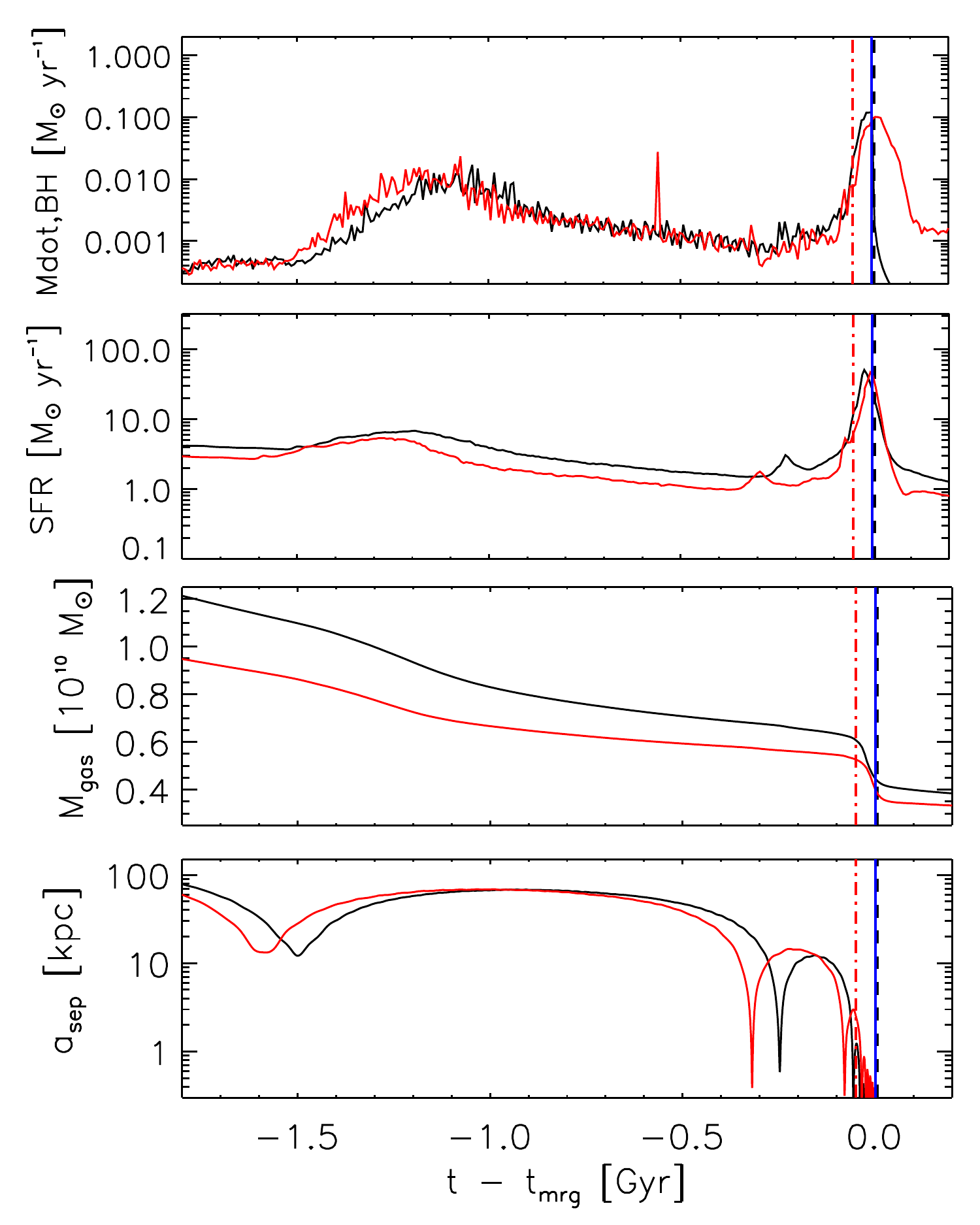}
\caption{Time evolution of relevant quantities throughout the galaxy merger for the recoiling SMBH (black curves) and dual SMBH (red curves) models. The quantity $t - $\tmrg\ is plotted on the $x$-axis; each model has slightly different \tmrg. From top to bottom, the panels show the total SMBH accretion rate, the global star formation rate (SFR), the total gas mass, and the 3-D SMBH separation prior to SMBH merger. In each panel, the dot-dashed red line denotes \tobs\ for the dual SMBH model, the black dashed line denotes \tobs\ for the recoiling SMBH model with \vk$ = 1428$ \kms, and the solid blue line denotes \tmrg. 
\label{fig:evol}}
\end{figure}

In order to compare our hydrodynamic models more directly with observations, we have employed the three-dimensional, polychromatic dust radiative transfer code \sunrise\ \citep{jonsso10}. The {\em HST}/ACS (F814W) image of CID-42 is shown in Figure \ref{fig:stars}a, along with \sunrise\ images in the same band for our GW recoil model (Figure \ref{fig:stars}b) and dual SMBH model (Figure \ref{fig:stars}b). Recall that in the GW recoil scenario, only one (post-merger) SMBH is present in CID-42, as opposed to the two SMBHs present in the latter model. Figures~\ref{fig:stars}b \& c show the mergers at the time of observation (\tobs), i.e. when the spatial offset (and for the recoil, the line-of-sight (LOS) velocity offset) of the SMBH(s) match the observed values. Although the recoiling SMBH position is marked in Figure \ref{fig:stars}b, the ejected-disk luminosity is not accounted for in the image, so there is no observable excess of emission from this point. In the dual SMBH model (Figure \ref{fig:stars}c), the SMBH positions correspond to the two central brightness peaks. 

The galactic morphology of each model is broadly similar to the observed morphology. While these models are not unique solutions, they have some generic properties. For example, the prominent asymmetric tidal feature can arise in the late stages of a major but unequal-mass merger, such as the $q=0.5$ mergers shown here. Fainter tidal features can also be seen in the lower-right (SW) edge of both the observed and simulated images. Some differences between the two models can be seen as well. The merger remnant with dual SMBHs appears to be somewhat ``rounder" than does the remnant with a recoiling SMBH. However, the bolometric luminosity of the recoil model ($L_{\rm bol} \sim 2.5 \times 10^{11}$ \lsun) is closer to that inferred for CID-42 from SED fitting ($L_{\rm bol} \sim 2.9\times10^{11}$ \lsun, FC10).  In the dual SMBH model, $L_{\rm bol}$  is only $\sim 6.5\times10^{10}$ \lsun. Table~\ref{table:params} summarizes the relevant physical parameters of our models in comparison to those inferred from observations of CID-42. 

Fig.~\ref{fig:evol} shows the merger evolution of various quantities for both the recoiling and dual SMBH models. Because the SFR of CID-42 inferred from SED fitting is $\sim 25$ \msunyr, the models require a starburst that more or less coincides with \tobs. Similarly, the estimated bolometric ($L_{24\mu{\rm m-UV}}$) luminosity of the SE source, assuming it is a Type 1 AGN, is $7.3\times 10^{10}$ \lsun, or $\sim 4\%$ L$_{\rm Edd}$ for an estimated SMBH mass of $6.5\times 10^7$ \msun\ (FC10). Thus, the SMBHs are also near their peak luminosity at the time of the SMBH merger. The presence of a stellar bulge in the progenitor galaxies acts to stabilize the gas and stellar disks against perturbations, such that the strongest gas inflow is delayed until the galaxies' final coalescence \citep[e.g.,][]{mihher96}. A high initial gas content in the progenitor galaxies (40-50\% of the disk mass) is also required for fueling rapid star formation and SMBH growth. 

The dual and recoiling SMBH models have very similar evolution, but because \tobs\ in the dual SMBH model necessarily occurs some time prior to the SMBH merger (\tmrg\ $-50$ Myr in our case), the system is observed before the peak SFR and $\dot M_{\rm BH}$. Specifically, the SFR at \tobs\ in the dual SMBH model is $\sim 6$ \msunyr, versus $\sim 21$ \msunyr\ for the recoil model. At \tobs, $\sim$ 50 -- 70\%  of the gas has been consumed. The remaining gas mass, $\sim 4 $ -- $5 \times 10^9$ \msun, might be detectable, and an estimate of its neutral fraction obtained, via CO or 21cm observations.

Our simulated merger remnants are consistent with being luminous infrared galaxies (LIRGs), as is CID-42. However, our \sunrise\ spectra are more IR-dominated than the observed SED; we find that at the time of observation, $L_{24\mu \rm m} / L_{\rm bol} \sim 0.2$ and 0.4 for the dual and recoiling SMBH models, respectively, compared with $L_{24\mu \rm m} / L_{\rm bol} \sim 0.1$ for the Seyfert 1.8 SED fit to CID-42 (FC10). In the recoiling SMBH case, this disparity owes partly to the fact that the post-recoil AGN luminosity is calculated analytically after the simulation, and thus does not contribute to the calculated SED. Similarly, in the dual SMBH case, the (smaller) SED mismatch arises partly because $L_{\rm bol}$ for the brighter AGN is a factor of $\sim 7$ lower than $L_{\rm bol,BH}$ inferred for CID-42 (FC10), such that both AGN contribute relatively little to the calculated SED. At the same time, the SFR at \tobs\ in the dual SMBH model is more than a factor of three lower than both the SFR in the recoil model and that inferred for CID-42 (FC12). However, the results presented here do not depend on the shape of the simulated SED. 

\setlength{\tabcolsep}{4pt}

\begin{table}
\begin{center}
\begin{tabular}{l l r r r r r}

& & \multicolumn{4}{c}{\bf Model} & \\
 \cline{3-6} \\[-6pt]
{\bf Quantity} & {\bf Units} &  {\bf high-\vk} & {\bf low-\vk} & \multicolumn{2}{r}{\bf dual SMBH} & {\bf CID-42} \\ \hline \\[-6pt]

\vk\ & [\kms] & 2470 & 1428 & \multicolumn{2}{r}{--} & -- \\[3pt]
$v_{\rm LOS}$/\vk\ & -- & 0.58 & 0.81 & \multicolumn{2}{r}{--} & -- \\[3pt]
$v_{\rm LOS}$(\tobs) & [\kms] & 1360 & 1180 & \multicolumn{2}{r}{--} & -- \\[3pt]
\tobs$-$\tmrg\ & [Myr] & $+1.3$ & $+5.9$ &  \multicolumn{2}{r}{$-50$} & -- \\[3pt]

$M_{\rm star}$ & [$10^{10}$ \msun] & 2.4 & 2.4 & \multicolumn{2}{r}{2.3} & 2.5 \\[3pt]
$M_{\rm gas}$ & [$10^{10}$ \msun] & 0.43 & 0.43 & \multicolumn{2}{r}{0.53} & -- \\[3pt]

SFR &  [\msunyr] & 21 & 21 & \multicolumn{2}{r}{6.3} & 25 \\[3pt]

$L_{\rm bol}$ & [$10^{10}$ \lsun] & 24.1 & 25.6 & \multicolumn{2}{r}{6.5} & 28.6 \\[3pt]
$L_{\rm 24 \mu m}$ & [$10^{10}$ \lsun] & 10.0 & 11.0 &  \multicolumn{2}{r}{1.4} & 3.2 \\[4pt]

& & & & {\bf \underline{BH1}} & {\bf \underline{BH2}} & {\bf \underline{SE}} \\[3pt]

$M_{\rm BH}$ & [$10^7$ \msun] & 0.98 & 0.97 & 0.45 & 0.21 & 6.5 \\[3pt]
$L_{\rm bol,BH}$ & [$10^{10}$ \lsun] & 0.50 & 0.32 & 0.12 & 0.89 & 7.3 

\end{tabular}
\end{center}
\caption{Relevant parameters of recoiling and dual SMBH models for CID-42, along with the parameters inferred from observations (FC10, FC12). \vk\ denotes the total kick velocity imparted to the merged SMBH for the recoil models; all other quantities are calculated at the time of observation (\tobs). The galactic $L_{\rm bol}$ and $L_{\rm 24\mu \rm m}$ for each model are obtained from radiative transfer calculations with SUNRISE. For the dual SMBH model, $M_{\rm BH}$ and $L_{\rm bol,BH}$ are given for each SMBH. The $M_{\rm BH}$ and $L_{\rm bol,BH}$ estimated for the SE optical source in CID-42 (FC10) are given in the last column. \label{table:params}}
\end{table}

\subsection{Recoiling SMBH Model}
\label{ssec:recoil_model}

\begin{figure}
\centering
\includegraphics[width=0.49\textwidth]{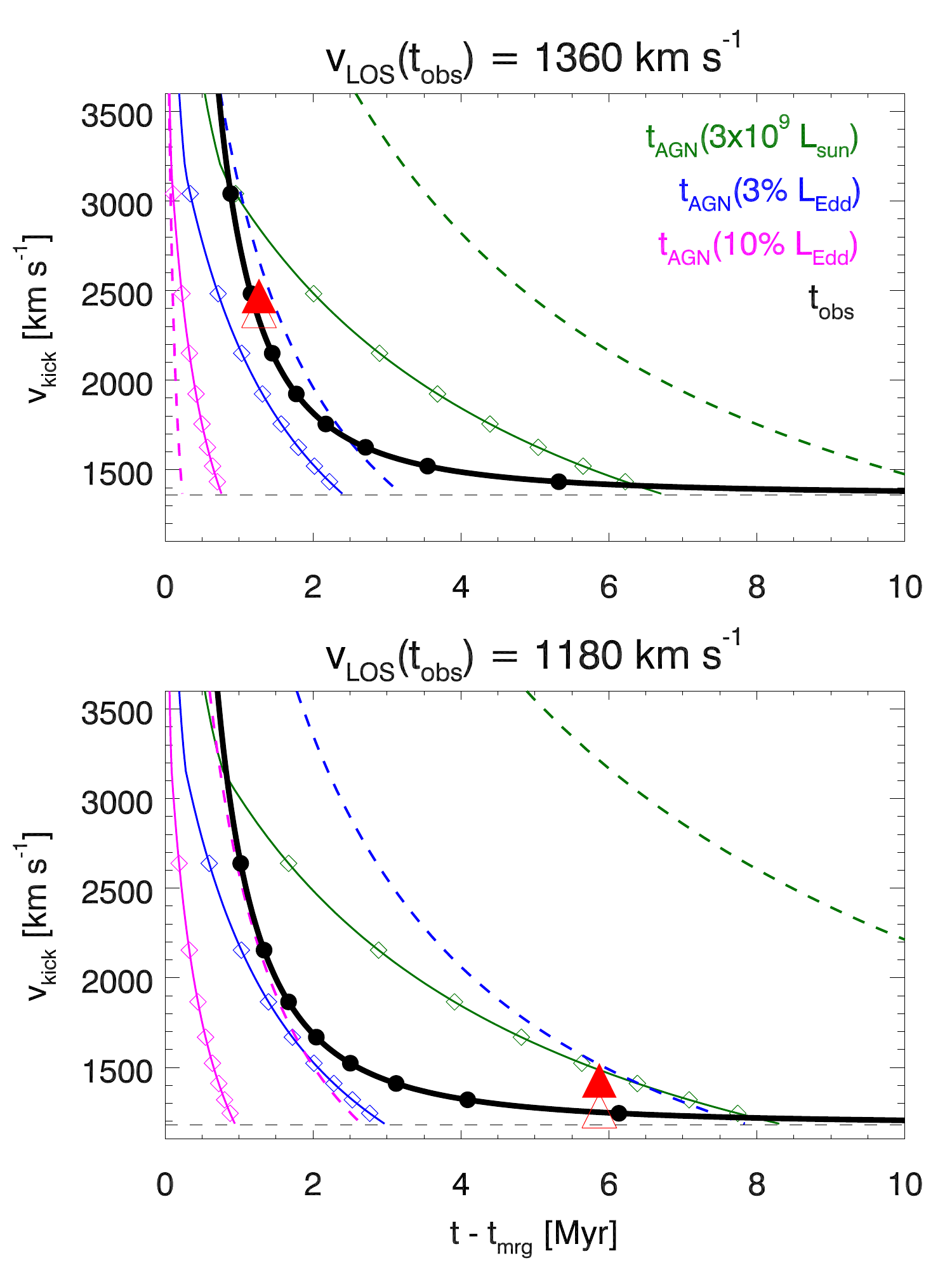}
\caption{\tobs, the post-recoil time at which the velocity and spatial offsets match the observations, and $t_{\rm AGN}$, the post-recoil AGN lifetime, are plotted for various total kick velocities (\vk) and LOS velocity components ($v_{\rm LOS}(t_{\rm mrg}) / v_{\rm k}$). Here, \tobs\ (solid black line) assumes no deceleration of the SMBH, a spatial offset of 2.46 kpc, and a line-of-sight velocity offset at \tobs, $v_{\rm LOS}(t_{\rm obs})$, of either 1360 \kms\ ({\em top panel}) or 1180 \kms\ ({\em bottom panel}), as discussed in the text. $v_{\rm LOS}(t_{\rm obs})$ is indicated by the dashed horizontal line in each plot. The solid black points indicate different values of $v_{\rm LOS}(t_{\rm mrg})/v_{\rm k}$, in intervals of 0.1. The colored solid lines correspond to \tagn\ for three definitions of the minimum observable AGN luminosity (see plot legend), using the values of $M_{\rm BH}$(\tmrg) and $\dot M_{\rm BH}$(\tmrg) calculated self-consistently from the merger simulations. Here too, values of $v_{\rm LOS}(t_{\rm mrg})/v_{\rm k}$ are indicated with open diamonds, in intervals of 0.1. The colored {\em dashed} lines show \tagn\ in a similar manner, except $M_{\rm BH}$(\tmrg) and $\dot M_{\rm BH}$(\tmrg) are calculated such that their values at \tobs\ match the observed values for CID-42. Note that while the brighter AGN phases are generally shorter than \tobs, the low-luminosity AGN lifetimes extend {\em beyond} \tobs\ for almost all \vk. Finally, the solid red triangle in each plot corresponds to the kick given, where \tobs\ assumes no deceleration. Because a small amount of deceleration occurs, the actual \tobs\ and $v(t_{\rm obs})$ are denoted by the open triangles. \label{fig:tscales}}
\end{figure}

\begin{figure}
\centering
\includegraphics[width=0.5\textwidth]{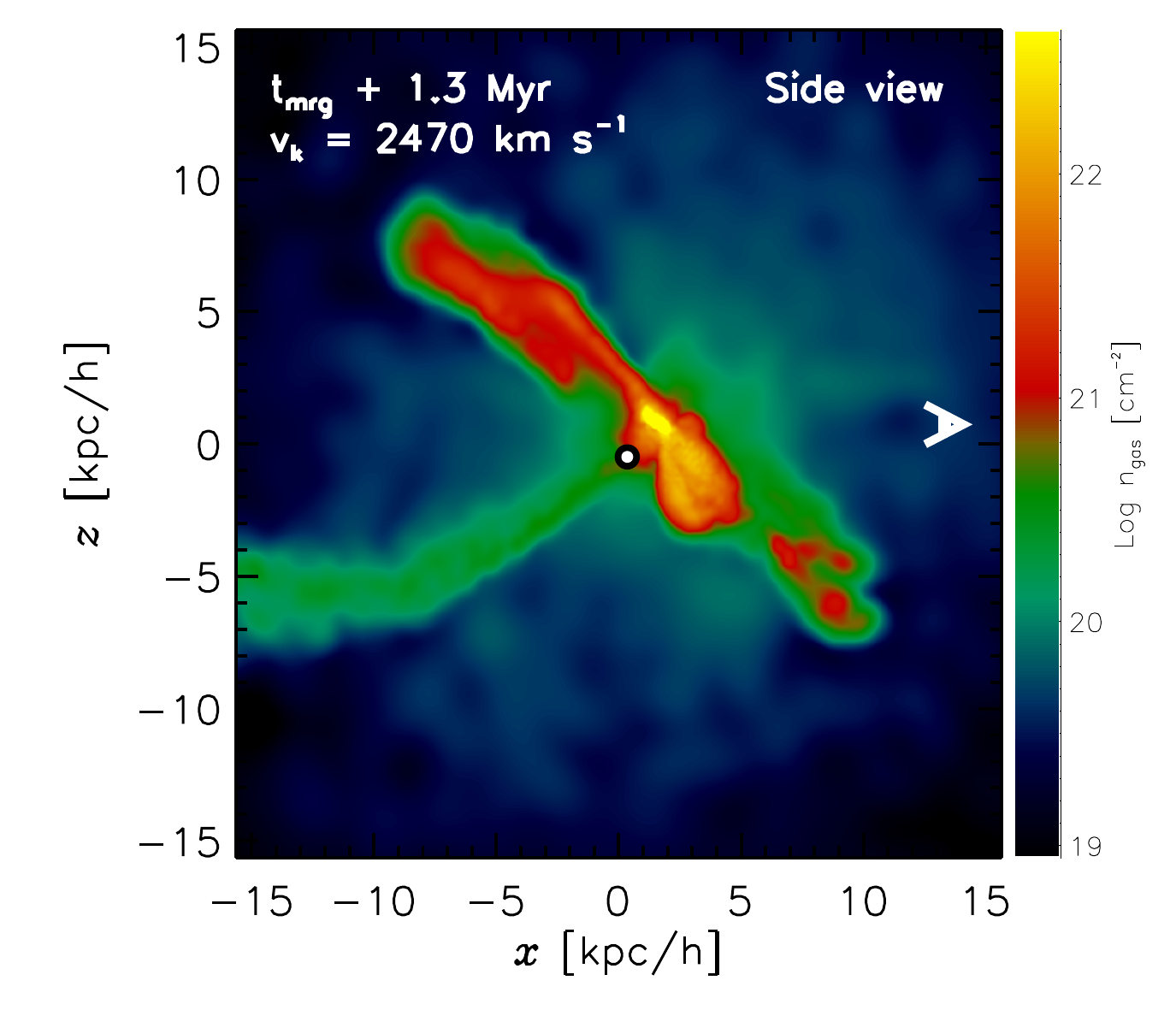}
\includegraphics[width=0.5\textwidth]{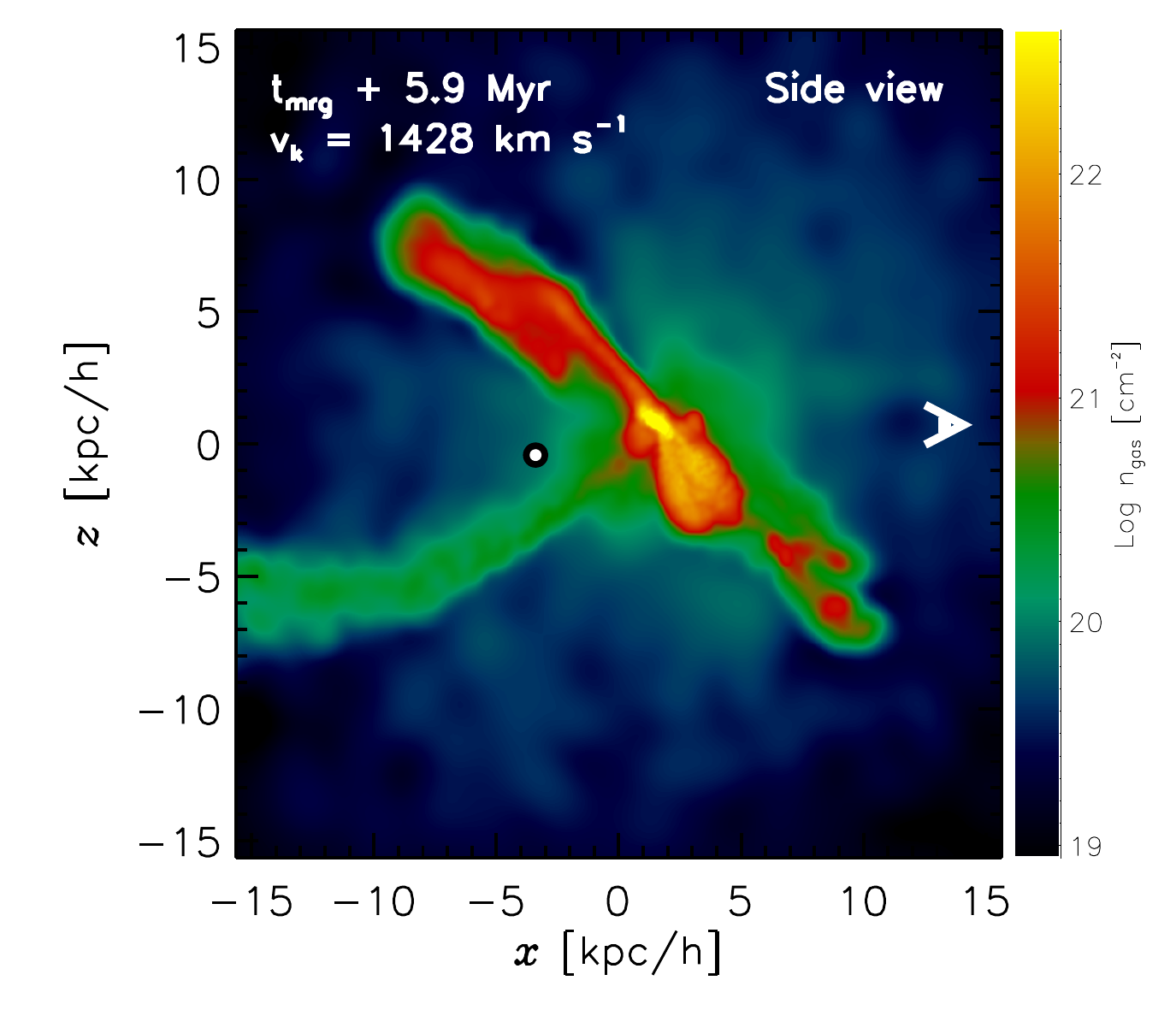}
\caption{Map of gas column density from side view for our best-fit recoiling SMBH models, shown at \tobs. In each case, the black \& white dot denotes the position of the BH, and the ``eye" symbol denotes the observer's LOS. The color scale shows the projected gas density in units of cm$^{-2}$. {\em Top panel}: GW recoil model with \vk\ $= 2470$ \kms and \tobs\ $= t_{\rm mrg} + 1.3$ Myr. {\em Bottom panel}: recoil model with \vk\ $= 1428$ \kms and \tobs\ $= t_{\rm mrg} + 5.9$ Myr. \label{fig:gas}}
\end{figure}

In our recoil models, the central galactic escape speed at the time of the SMBH merger is 773 \kms. Thus, we expect \vk\ $\gg$ \vesc\ for a recoil kick of $\sim 1300$ \kms\ in CID-42, such that the recoil event must have occurred recently. In order to match the position of the SE relative to the NW source, we direct the kick at an angle of $225^\circ$ in the plane of the sky. The kick velocity then has two free parameters: the fraction of the total kick directed in the line of sight, and the amount of the measured 1360 \kms\ BL/NL offset that is due to the motion of the SMBH. Specifically, if the NL emission originates from the NW source and the BL emission from the SE, then a spatial-offset correction must be applied to the measured  velocity offset, such that the actual BL/NL velocity offset would be 1180 \kms\ (FC10). 

We consider models for two different kick speeds. {\rev The ``low" kick, $v_{\rm k} = 1428$ \kms, corresponds to $(v_{\rm LOS}(t_{\rm mrg}) / v_{\rm k})^2 = 0.9$ and $v_{\rm LOS}(t_{\rm obs}) = 1180$ \kms, and the ``high" kick, \vk\ $= 2470$ \kms, corresponds to $(v_{\rm LOS}(t_{\rm mrg})/v_{\rm k})^2 = 1/3$ and $v_{\rm LOS}(t_{\rm obs}) = 1360$ \kms\ (see Table~\ref{table:params}).} Note that although \vk\ $\gg$ \vesc, there is a small difference between $v_{\rm LOS}(t_{\rm mrg})$ and $v_{\rm LOS}(t_{\rm obs})$ owing to deceleration. We choose these values of \vk\ because velocities lower than $\sim 1400$ \kms\ would require the kick to be oriented almost entirely along the LOS, and velocities much higher than $\sim 2500$ \kms\ would have a very small probability of occurring. These probabilities are discussed further in \S~\ref{ssec:recoil}, but we note that both the ``low" and ``high" kick velocities are quite large in absolute terms. 

For the low kick, \tobs\ occurs 5.9 Myr after the recoil. The mass of the ejected gas disk is $\sim 1.4\%$ $M_{\rm BH}$, and $L_{\rm bol,BH} = 56\% L_{\rm Edd}$ at the time of the kick. At \tobs, the AGN luminosity has declined to 1\% $L_{\rm Edd}$, or $3.2 \times 10^{9}$ \lsun\ (assuming a radiative efficiency of 0.1). For the high kick, a gas mass of only $\sim 0.5\%$ $M_{\rm BH}$ remains bound to the recoiling SMBH, but because the system would be observed sooner (\tobs\ $=$ \tmrg\ $+ 1.3$ Myr), the AGN luminosity is slightly higher.  These values are summarized in Table~\ref{table:params}. While these AGN luminosities are lower than that estimated for CID-42 via SED fitting (FC10), the  Eddington ratios of the order of $\sim 1\%$ are consistent.

The relationships between \tobs, $t_{\rm AGN}$, and \vk\ are illustrated in Fig.~\ref{fig:tscales}. For the low-luminosity AGN definition ($L_{\rm bol,BH} > 3\times10^{9}$ \lsun), the AGN lifetimes of the recoiling BHs (as defined in Eqn.~\ref{eqn:tagn}) are 2.3 Myr and 6.4 Myr for the low and high kicks, respectively; i.e., both have $t_{\rm AGN} > $ \tobs. In fact, for the observed parameters of CID-42, virtually {\em all} allowed values of \vk\ and \tobs\ result in a low-luminosity recoiling AGN with $t_{\rm AGN} > $ \tobs.

Moreover, we expect the recoiling AGN lifetimes to be even longer than these estimates, because the AGN luminosities and SMBH masses calculated self-consistently from our recoil simulations are lower than the values inferred for the SE source. We can use the fully-analytic ejected disk model to calculate the post-recoil AGN luminosities and lifetimes for arbitrary values of $M_{\rm BH}$(\tmrg) and $\dot M_{\rm BH}$(\tmrg). Specifically, we can match $\dot M_{\rm BH}$(\tobs) to the observed luminosity of the SE source in CID-42 and extrapolate backward to obtain its value at \tmrg\ for each recoil model. The estimates of \tagn\ using these values (along with $M_{\rm BH}$ inferred for the SE source) are indicated by the dashed lines in Figure~\ref{fig:tscales}. Here, the SMBH in each recoil model has $L_{\rm bol,BH}/L_{\rm Edd} > 3\%$ at \tobs. By the lowest-luminosity AGN definition, \tagn\ is 4.9 Myr and 18 Myr for the low- and high-kick models, respectively. Therefore, the fuel supply carried along with the ejected SMBH should not be a limiting factor for the recoiling AGN scenario. 

Another key consideration is whether the recoiling AGN is able to ionize gas and produce NL emission as it leaves the galaxy. Based on the current data, it is unclear whether the observed NL emission is coming from the SE or NW source. However, this will be constrained in the near future via individual {\em HST} spectra of each source (Civano et al. 2012c, in preparation), so it is important to know whether a recoiling AGN might be capable of producing the observed emission lines. 

It is clear from Figure \ref{fig:gas} that the recoiling SMBH is ejected to a low-gas-density region within at most a few Myr. In both recoil models, all of the gas near the SMBH at \tobs\ is warm and diffuse, with an average temperature of a few $\times 10^5$ K and densities of $10^{-1.5}$ and $10^{-2.5}$ cm$^{-3}$ for the high- and low-kick models, respectively. None of this gas is condensed into the ``cold" phase of the multiphase model for the interstellar medium (ISM) used in \gadthree\ \citep{sprher03}. Thus, we would not {\em a priori} expect any cold clouds to be {\em locally} ionized by the recoiling SMBH. However, sufficiently luminous AGN may ionize gas from outside the galaxy, similar to what is observed in some merging galaxies containing quasars \citep[e.g.,][]{dasilv11}. 

\citet{blecha12} have developed a semi-analytic model for estimating the narrow-line emission around AGN in hydrodynamic simulations of galaxy mergers. We have applied this model to our simulations of recoiling SMBHs in CID-42, including accretion from the ejected gas disk. In order to estimate the lifetime of observable NL emission, we assume a ratio of the \oiii\ $\lambda$5007 luminosity to the bolometric AGN luminosity of $L_{\rm [OIII]}/L_{\rm bol,BH} = 10^{-3}$. This is the ratio of the total \oiii\ luminosity to the luminosity of the SE source estimated by FC10. In reality, this ratio may be higher (such that the estimated NL lifetime would be an upper limit), because $L_{\rm [OIII]}$ is measured from the total continuum rather than just the AGN continuum. However, additional uncertainty comes from the fact that $L_{\rm [OIII]}$ is measured from a slit spectrum, whereas $L_{\rm bol,BH}$ is measured in a circular aperture; the direction of this effect is less clear. We therefore consider the ratio $10^{-3}$ to be a reasonable guess, but not necessarily an upper limit. 

The NL model of \citet{blecha12} calculates the \hbeta\ rather than the \oiii\ luminosity, because the former is less sensitive to the conditions in the ISM. Thus, we must also assume an \oiii/\hbeta\ ratio. The ratio estimated from the spectra of CID-42 is $\sim$ 6 (JC09, FC10), so we use a ratio of 10 as a conservative upper limit. This implies a minimum observable $L_{\rm H\beta}/L_{\rm bol,BH}$ of $10^{-4}$. 

In this analysis of the NLR, we must account for the difference between the AGN luminosities calculated self-consistently in our simulations and the AGN luminosity of the SE source estimated from its F814W magnitude. Taking the values from the simulation, we find that the recoiling AGN can produce NL emission above the estimated minimum value of $3-5 \times 10^5$ \lsun\ for only about $1-2$ Myr after the kick, depending on \vk. In each case the NL lifetime is shorter than \tobs. {\rev Note that any observed photoionization of NLRs must be ongoing; otherwise, NL AGN emission would essentially last for only a light crossing time ($\la 10^3$ yr for a typical NLR). After this, \oiii\ would disappear on a timescale of $\sim$ years via charge-transfer recombinations (destroying the large \oiii/\hbeta\ ratio characteristic of AGN), followed by H recombination on a timescale of $\la 10^3$ yr.}

If instead we assume the recoiling AGN has $M_{\rm BH}$(\tmrg)$ = 6.5\times10^7$ \msun\ and $L_{\rm bol,BH} = 7.3\times10^{10}$ \lsun\ (the estimates for the SE source in CID-42 assuming it is a Type 1 AGN, FC10), the NL lifetimes are longer, despite the stricter minimum \hbeta\ luminosity of $7.3 \times 10^6$ \lsun. Specifically, the NL lifetime for the low kick is 4.0 Myr. This is still shorter than \tobs\ $= 5.9$ Myr, so the recoiling AGN in this model should not produce observable NL emission. In the high-kick model, however,  the NL lifetime is 1.6 Myr, compared to \tobs\ $= 1.3$ Myr. This suggests that a recoiling AGN in CID-42 could to produce the observed narrow lines, provided the total kick velocity is sufficiently large ($\ga 2000$ \kms).

If the NL emission arises from stellar photoionization in the NW source, the recoiling AGN scenario does not require that the kicked SMBH produce observable NLs. However, the current spectra of CID-42 have AGN-like narrow-line flux ratios (JC09). This suggests that the recoil scenario requires either a large total kick velocity or an unusual star-forming region in the NW source. Alternatively, a recoiling AGN might be able to produce NL emission more locally if the SMBH happened to be ejected along a dense tidal stream or if the galactic line-of-sight gas distribution differed substantially from that of our model. 

Finally, we note that the total LOS gas column density to the position of the BHs is low; log $N_{\rm gas} \approx 20.8 - 20.9$ cm$^{-2}$. Thus, attenuation is insignificant along the LOS to the recoiling AGN, even though it is recoiling {\em away} from the observer (as inferred from the redshifted broad \hbeta\ line). 

\subsection{Dual SMBH Model}
\label{ssec:dual_model}

In the dual SMBH model, the projected SMBH separation of 2.5 kpc occurs at \tobs\ $=$ \tmrg\ $-50$ Myr. The slightly less dissipative merger in this model is key to reproducing the observed morphology prior to the SMBH merger. If we simply examine the pre-merger snapshot from the recoil model in which the SMBHs have a projected separation of 2.5 kpc, we find that the galaxy is substantially more disturbed, with multiple prominent tidal features and a more compact and asymmetric core. 

The SMBH masses at \tobs\ are $4.5\times10^6$ \msun\ and $2.1\times10^6$ \msun. Thus, each SMBH is more than a factor of ten smaller than the estimated mass of the SE source in CID-42. The less massive SMBH is actually {\em more} luminous, with $\dot M = 13\% \dot M_{\rm Edd}$ and $L_{\rm bol,BH} = 8.9\times 10^{9}$ \lsun. The more massive SMBH is accreting at only 0.8\% $\dot M_{\rm Edd}$, which corresponds to a bolometric luminosity $L_{\rm bol,BH} = 1.2\times 10^{9}$ \lsun. Notably, this lower-luminosity AGN is consistent with the upper limit on the X-ray luminosity for the NW source of CID-42 ($L_{\rm X} < 10^{42}$ erg s\inv), so this alone does not place constraints on whether it may contain a SMBH.

However, because the more luminous AGN in our model is about eight times fainter than the luminosity of the SE source in CID-42, the AGN luminosity {\em ratio} in our model may be more informative than the absolute luminosities. In CID-42, the measured F814W luminosity ratio of the NW and SE sources is $\sim 2.5$, and the NW source is brighter (FC10). Given the uncertainty as to whether an AGN exists at all in the NW source, this tells us little about the possible {\em AGN} luminosity ratio. Instead, we compare the limit on X-ray emission from the NW source with the bolometric luminosity ratio of our simulated AGN. Luminosity ratios greater than 25 correspond to the 4\% upper limit on the fraction of hard X-ray flux originating in the NW source (FC12). In the kpc-scale phase of our simulation (defined loosely as SMBH separations $< 10$ kpc), the AGN luminosity ratio is above 25 for $\la 1\%$ of the time, and not at all for separations $< 5$ kpc. At \tobs, the luminosity ratio of the two AGN is 7.5. This suggests that an AGN pair with an {\em intrinsically} large luminosity ratio is a possible but relatively unlikely explanation for the lack of X-ray emission from the NW optical source of CID-42. 

We must also consider that a highly {\em obscured} AGN could be present in the NW source. Given the lack of detection of X-rays from this location, an obscured AGN here would have to be Compton thick (log $N_{\rm H} > 24$ cm$^{-2}$). The total LOS gas column densities to the SMBHs in our model are log $N_{\rm gas} \approx 23.8$ \& 22.1 cm$^{-2}$ for the more and less luminous AGN, respectively. Notably, the column density of the more luminous AGN is near the Compton-thick limit, though these numbers should not be taken at face value. The Compton-thick limit refers to the equivalent neutral hydrogen column density of the gas, which depends on the ionic fraction of each element, whereas we obtain the total gas column density from our simulations. Our column densities therefore represent an upper limit on $N_{\rm H}$. Further, the fact that the more intrinsically luminous AGN is hosted in the more obscured density cusp works against the need for a large emergent luminosity ratio in the hard X-ray band. 

Although we do not find strong evidence for an AGN obscured by the galactic environment of CID-42, many AGN are thought to contain dense, obscuring material on small scales, well below the resolution limit of our simulations. If Compton-thick clouds fell along our line of sight to an AGN in the NW source, this could cause significant attenuation through the hard X-ray band. In this case, much of the AGN luminosity would be reprocessed as IR emission, creating an IR excess relative to that expected for a star-forming galaxy with a SFR of $\sim 25$ \msunyr. Thus, the fact that the SED of CID-42 is well-fit by a template for a star-forming galaxy (FC10) argues against the presence of a Compton-thick AGN.

Strong Fe K$\alpha$ emission \citep[EW $\sim$ 1--2 keV][]{matt03} is also expected in heavily obscured X-ray sources, (those with log $N_{\rm H} \ga 23$ cm$^{-2}$). A Fe K$\alpha$ emission feature with EW $\sim$ 100--500 eV is indeed observed in both the {\em Chandra} and {\em XMM-Newton} data for CID-42 (FC10). However, because this equivalent width could easily be produced by a Type 1 AGN in the SE source, we cannot determine whether the Fe K$\alpha$ line originates from an obscured AGN in the NW source.

As with the recoil scenario, we have applied the NL model of \citet{blecha12} to our dual SMBH simulation. In this case, we wish to know whether the system should appear as a NL AGN and whether the emission lines may be double-peaked, as indicated by the spectra in JC09. We find that at \tobs, the lower-luminosity AGN has a narrow-\hbeta\ luminosity too low to be observable. The more luminous AGN has a higher ionizing photon flux, but as it also resides in a higher-density region than the first AGN, its ionization parameter is marginally too low by the criteria of \citet{blecha12}. The lack of clear evidence for strong NL emission from these AGN reflects their low luminosities relative to the SE source of CID-42. Regardless of the NL luminosities, their relative LOS velocity in this snapshot is only 64 \kms. Because this is smaller than the typical FWHM of NLs, it is unlikely that double-peaked emission lines would be resolvable in this case. Our model for CID-42 is not a unique solution, however, and the LOS velocities for projected SMBH separations of 2 -- 3 kpc range from $\sim 0 - 150$ \kms. Accordingly, we cannot rule out the possibility that a system like CID-42 could produce double-peaked NLs.  Civano et al. (2012c, in preparation) have obtained a new DEIMOS spectrum of CID-42 with much higher S/N, which will reveal definitively whether the NLs are double-peaked.

\section{Discussion}
\label{sec:discuss}

\subsection{GW Recoil Scenario}
\label{ssec:recoil}

Several observed features of CID-42 support the scenario that it contains a single, rapidly-recoiling SMBH. The offset between the broad and narrow emission lines is expected to be characteristic of recoiling AGN \citep{merrit06}. The single X-ray source coincident with the SE optical brightness peak is also consistent with the GW recoil scenario, in which the SE source corresponds to a recoiling AGN that has left the central stellar cusp behind (FC12). Additionally, the best-fit surface brightness decomposition performed by FC10 consists of an extended (Sersic) component for the NW source and a point-like component for the SE source, consistent with a stellar cusp and a recoiling AGN, respectively.

{\rev We note that because CID-42 is actively star-forming, implying a high gas content, we expect the merger remnant to have a stellar cusp even after the displacement of stars via binary SMBH inspiral \citep[e.g.,][]{milmer01,yu02,merrit06} and the recoil event itself \citep{boylan04,guamer08}. Gas-rich major mergers trigger a rapid inflow of cold gas and subsequent star formation \citep[e.g.,][]{barher96,mihher96,mayer07,hopkin08d}, thereby maintaining a steep central density profile. Stellar interactions with the SMBHs should be relatively unimportant in such cases, because hydrodynamic processes drive the efficient formation and inspiral of the binary SMBH \citep{mayer07}. Furthermore, \citet{blecha11} found that recoil kicks may actually enhance the central stellar density by displacing AGN feedback and delaying the quenching of star formation. We accordingly refer to a stellar ``cusp" when discussing the NW source, though the observations do not exclude the presence of a stellar core on scales below the {\em HST} image resolution of 0.03" (150 pc).}

Our recoiling SMBH models of CID-42 are able to reproduce the morphology and stellar mass of CID-42 in the late stages of a gas-rich, major galaxy merger. The peak SFR and $\dot M_{\rm BH}$ occur near the time of SMBH merger, and their values at \tobs\ are consistent with estimates from the observations (SFR $\sim 25$ \msunyr\ \& $L_{\rm bol,BH}/L_{\rm Edd} \sim$ few percent; FC10, FC12). 

Less clear is whether a recoiling AGN could produce the observed NL emission in CID-42. If the total kick velocity is large, such that the system is observed soon after the kick, the recoiling AGN may still be close enough to the dense gaseous region to produce observable NL emission. Specifically, we find that the recoiling AGN in our high-kick model (\vk\ $= 2470$ \kms) could produce the observed NLs in CID-42, but that it probably could not do so with a lower kick of 1428 \kms.

In the latter case, special configurations may be required for the recoiling AGN to produce the NL emission, such as ejection of the SMBH along a dense tidal stream. If the observed NLs originate from star formation in the NW source rather than from the recoiling SMBH, then the recoil scenario requires unusual line flux ratios for a star-forming region. The origin of the NL emission is therefore an important consideration in evaluating the likelihood that CID-42 contains a recoiling AGN, and underscores the need for corroborating evidence. Individual spectra of the NW and SE sources will provide important constraints in this regard (Civano et al. 2012c, in preparation), and IFU data could also provide useful information about the spatial distribution of narrow (and broad) emission line regions in CID-42.

IFU spectra, or HST imaging in multiple bands, would also constrain the colors of the NW and SE sources, and of the galaxy as a whole. Specifically, this would indicate whether the galaxy is indeed blue and actively star-forming, as predicted by SED fitting (FC10, FC12) and our simulations. Such observations would also reveal whether the starburst is concentrated within the NW source, and they would provide an independent check on the starburst age of a few Myr estimated from SED fitting (FC12). 

Another consideration is that a merger remnant with a recoiling AGN should contain, in principle, one central brightness peak (the stellar cusp) and one offset brightness peak (the AGN). An AGN pair, in contrast, should have both brightness peaks offset from the center. However, this is an inconclusive diagnostic for CID-42, because its center of stellar light is not well constrained. The surface brightness decomposition of FC10 places the center between the NW and SE sources, but it is consistent with being at the position of the NW source within the errors. The central stellar cusp could also be intrinsically asymmetric in the unrelaxed merger remnant, and tidal features seen in projection could contribute light to the apparent galactic disk. 

As noted in \S~\ref{ssec:recoil_model}, even the lowest possible recoil velocity for CID-42 is quite large ($\ga 1300$ \kms). The intrinsic probability of such a large kick is not clear. Recoil kick probability distributions have been derived from numerical relativity simulations \citep[e.g.,][]{baker08,vanmet10,lousto12}. However, the distributions of progenitor mass ratios, spins, and spin orientations are still quite uncertain. In particular, if a circumbinary gas disk forms around the inspiraling SMBHs, it may align or partially align their spins \citep{bogdan07,dotti10}.  The maximum recoil velocity resulting from the merger of BHs with aligned spins is only $\sim 200$ \kms.  Partial alignment may also occur via GR precession effects \citep{kesden10,berti12}, though this mechanism can also result in {\em anti}-alignment of the BH spins, yielding substantially larger kicks. At any rate, it is clear that the GW recoil scenario for CID-42 requires that spin alignment prior to BH merger is inefficient, at least in some cases. 

\subsection{Gravitational Slingshot Recoil Scenario}
\label{ssec:slingshot}

A rapidly-recoiling SMBH could also result from the gravitational slingshot mechanism. As described in \S~\ref{sec:intro}, this requires a long binary SMBH inspiral time, such that a subsequent galaxy merger introduces a third SMBH before the initial SMBH binary has coalesced. A three-body encounter may result in the ejection of the lightest SMBH at high speeds, and may also cause the rapid merger of the remaining SMBHs \citep{hofloe07}. Thus, in the slingshot-recoil scenario, two {\em or three} SMBHs would be present in CID-42: the SE source would be the ejected SMBH, and the NW source would contain the SMBH(s) left behind. 

While we have not constructed a model for this scenario, much of our analysis for the GW recoil model can be applied here. If the ejected SMBH retains a gas disk after the three-body encounter (such that it appears as an AGN), the accretion rate will evolve as described in \S~\ref{ssec:sims}. The relationship between \vk\ and \tobs\ would also be the same as in Figure \ref{fig:tscales}. If the NW source contains an obscured, radio-loud AGN, this could be detected with very long baseline interferometry (VLBI). If it contains a {\em binary} AGN that survived the three-body encounter, the binary might be spatially resolvable with VLBI imaging, provided its separation were at least a few parsec and the AGN were both radio loud. However, binary radio-detected AGN are intrinsically quite rare, with only one example known \citep{rodrig06} despite a dedicated search of thousands of archival VLBI images \citep{burkes11}.

If SMBH binaries do not merge efficiently, the probability of such triple SMBH systems forming is non-negligible, especially at high redshift \citep[$z \ga 2$,][]{kulloe12}. In our simulations, efficient merging of the SMBHs is merely an assumption owing to resolution limits. However, smaller-scale simulations of inspiraling SMBHs indicate that gas drag \citep[e.g., ][]{escala05,dotti07} or non-axisymmetry of the stellar potential \citep[e.g., ][]{berczi06,khan12} can prevent SMBH binaries from ``stalling" \citep[the so-called ``final parsec problem", e.g.,][]{begelm80,yu02}. CID-42 is clearly a gas-rich system with a disturbed stellar distribution. Thus, while the possibility of gravitational slingshot recoil cannot be excluded, we consider it to be unlikely. Put another way, if new evidence points toward a slingshot recoil in CID-42 (for example, if compact radio emission were detected in the NW source and consistent velocity offsets were found in multiple broad lines in the SE source), the implications for binary SMBHs would be far-reaching. This would indicate that many unmerged SMBH binaries may be lurking below resolution limits even in gas-rich environments, to the detriment of SMBH merger event rates for future gravitational-wave observatories.

\subsection{Dual SMBH Scenario}
\label{ssec:dual}

The observations of CID-42 are also consistent with an SMBH pair rather than a recoiling SMBH. In this scenario, the NW and SE sources are two inspiraling SMBHs with a 2.5 kpc separation. However, the $\sim 1300$ \kms\ offset of the broad \hbeta\ line is too large to be associated with orbital motion on kiloparsec scales. Many AGN spectra show offset broad lines that are often attributed to outflows, but the offset line in CID-42 is atypically large. CID-42 constitutes 1/323, or {\rev 0.3\%}, of BL AGN in the {\em Chandra}-COSMOS survey with spectroscopic redshifts \citep{elvis09}, and less than $1$\% of SDSS quasars have broad \hbeta\ lines offset by $> 1000$ \kms\ \citep{bonshi07,eracle12}.

Theoretically, a projected SMBH separation of 2.5 kpc could also occur during an early close passage of the merging galaxies, or even during a flyby encounter. However, this is an unlikely explanation for CID-42. Its tidal features are indicative of a late-stage merger, and the highest $\dot M_{\rm BH}$ and SFR typically occur near coalescence (see Fig.~\ref{fig:evol}). An unbound encounter would be required to explain the 1300 \kms\ BL offset via relative galactic motion, whereas the maximal relative speeds in our galaxy merger simulations are $\sim 500 - 600$ \kms. In such a flyby, only very specific configurations and timing would allow a projected separation of 2.5 kpc to be observed. 

Our dual SMBH model is well-matched to the morphology and stellar mass of CID-42, but a slightly less dissipative merger (lower \fgas) is required in order to match the observed morphology prior to the SMBH merger. \tobs\ also occurs slightly before the peak SFR and SMBH accretion rate. As a consequence, the SFR, $\dot M_{\rm BH}$, and luminosities are lower in the dual SMBH model than the estimates for CID-42. 

Owing to these low AGN luminosities of a few $\times 10^{9}$ \lsun, their NL emission \citep[as determined from the model of][]{blecha12} may be too faint to be observable. Also, their relative LOS velocity of only 64 \kms\ is unlikely to be spectroscopically resolvable. This is consistent with the supposition that the double-peaked NL structure in the DEIMOS spectrum (JC09) does not arise from dual SMBH orbital motion.

The non-detection of X-ray emission in the NW source (FC12) places some constraints on whether it may host an AGN. The NW source could contain a {\em quiescent} SMBH with $L_{\rm X} < 10^{42}$ erg s\inv, but our models indicate that two SMBHs in a late-stage, gas-rich major merger are unlikely to have a large intrinsic luminosity ratio. Specifically, an X-ray faint AGN in the NW source requires a luminosity ratio $> 25$, which occurs for $< 1\%$ of the kpc-scale phase in the dual SMBH simulation.

The NW source could also contain a Compton-thick AGN.  The upper limits on the gas column densities along the LOS to each SMBH (log $N_{\rm gas} \approx 22.1$ \& 23.8 cm$^{-2}$) indicate that these AGN should be absorbed and highly obscured, respectively, but not Compton-thick. The hard X-ray detected AGN pair in NGC 3393, for example, has estimated line-of-sight column densities of a few $\times 10^{23}$ cm$^{-2}$ \citep{fabbia11}. It is possible that obscuration on sub-resolution scales could hide an AGN in the NW source, but in this case, we expect the SED to be more IR-dominated than what is observed. Finally, while a strong iron emission line can indicate the presence of a heavily obscured AGN, this diagnostic is not informative for CID-42. A fairly strong Fe K$\alpha$ emission line is detected, but it is impossible to say how much, if any, of this emission is contributed by the NW source, which has a weaker continuum by a factor of at least 25.  

The detection of compact radio emission from either or both optical brightness peaks in CID-42 would provide strong evidence for the presence of one or two AGN. As mentioned above, however, radio-detected AGN {\em pairs} appear to be very rare, with only one radio-loud kpc-scale quasar pair \citep[3C 75; ][]{owen85} and one pc-scale AGN pair \citep{rodrig06} known. Regardless, given the ambiguous nature of the NW optical source and the X-ray non-detection there, the possibility of a radio detection is enticing, as this would provide the clearest evidence for an obscured AGN in the NW source. 

\subsection{Summary}
\label{ssec:summary}

We find that neither the recoiling nor dual SMBH scenarios for CID-42 can be ruled out based on current data. It is clear that whatever its true nature, CID-42 is not a ``typical" galaxy even among the rather exotic subsets that host dual or recoiling AGN. Both scenarios require some extraordinary features, such as a dual AGN with an unusually large broad line offset or a rapidly-recoiling AGN with an atypical narrow line region. Through our analysis we have identified and constrained the requisite properties of CID-42 in either case. These constraints combined with further observations in the coming months and years, such as spatially-resolved and higher-quality optical spectra, radio imaging, and CO or 21 cm observations, should rapidly narrow the possible origins of this intriguing system.

%\vspace{-8pt}

\section*{acknowledgments}
We would like to thank Julie Comerford for useful discussions and Chris Hayward, Patrik Jonsson, and Greg Snyder for helpful guidance with using \sunrise. This work was supported in part by NSF grant AST-0907890 and NASA grants
NNX08AL43G and NNA09DB30A.

%\vspace{-10pt}

\bibliography{refs_cid42_rev3_UPDATED}

\end{document}